\documentstyle[aps,prd,floats]{revtex}
\input{psfig.tex}

\begin{document}
\draft
\preprint{\vbox{\hbox{CU-TP-719}
                \hbox{CAL-589}
                \hbox{SU-4240-622}
                \hbox{CfA-4237}
}}

\title{Cosmological-Parameter Determination with Microwave
Background Maps}

\author{Gerard Jungman\footnote{jungman@npac.syr.edu}}
\address{Department of Physics, Syracuse University,
Syracuse, New York~~13244}
\author{Marc Kamionkowski\footnote{kamion@phys.columbia.edu}}
\address{Department of Physics, Columbia University,
New York, New York~~10027}
\author{Arthur Kosowsky\footnote{akosowsky@cfa.harvard.edu}}
\address{Harvard-Smithsonian Center for Astrophysics,
60 Garden Street, Cambridge, Massachusetts~~02138
\\and\\
Department of Physics, Lyman Laboratory, Harvard University,
Cambridge, Massachusetts~~02138}
\author{David N.~Spergel\footnote{dns@astro.princeton.edu}}
\address{Department of Astrophysical Sciences, Princeton University,
Princeton, New Jersey~~08544
\\and\\
Department of Astronomy, University of Maryland, College Park,
Maryland 20742}

\date{December 1995}
\maketitle

\begin{abstract}
The angular power spectrum of the cosmic microwave
background (CMB) contains information on virtually all cosmological
parameters of interest, including the geometry of the 
Universe ($\Omega$), the baryon density,
the Hubble constant ($h$), the cosmological constant
($\Lambda$), the number of light neutrinos, the ionization
history, and the amplitudes and spectral indices of the
primordial scalar and tensor perturbation spectra.
We review the imprint of each
parameter on the CMB.  Assuming only that the primordial
perturbations were adiabatic, we use a covariance-matrix
approach to estimate the precision with
which these parameters can be determined by a CMB temperature map as 
a function of the fraction of sky mapped, the level of pixel
noise, and the angular resolution.
For example, with no prior information about any of the
cosmological parameters, a full-sky CMB map with $0.5^\circ$
angular resolution
and a noise level of 15 $\mu$K per pixel can determine $\Omega$,
$h$, and $\Lambda$ with standard errors of $\pm0.1$ or better, and
provide determinations of other parameters which are
inaccessible with traditional observations.  Smaller beam
sizes or prior information
on some of the other parameters from other observations improves
the sensitivity.  The dependence on the the underlying cosmological
model is discussed.
\end{abstract}


\section{INTRODUCTION}

One of the fundamental goals of observational cosmology today is
measurement of the classical cosmological parameters:
the total density (or equivalently, the geometry) of the Universe,
$\Omega$; the cosmological constant $\Lambda$; the baryon
density $\Omega_b$; and the Hubble constant $H_0$.  Accurate
measurement of these quantities will test the
cornerstones of the hot big-bang theory and will provide answers
to some of the
outstanding questions in cosmology.  For example, determination
of the geometry of the Universe will tell us the ultimate fate
of the Universe and test the inflationary paradigm, while
an independent check of $\Omega_b$ can confirm
the predictions of big-bang nucleosynthesis.

In addition, parameters describing primordial perturbations
are related to the origin of large-scale structure in the Universe
and may shed light on a possible inflationary epoch.
Perhaps the most important of these are the
normalization $Q_S$ and spectral index $n_S$ of the primordial
spectrum of scalar perturbations that gave rise to the observed
structure.  Inflation may produce a spectrum of 
gravity waves, quantified by an amplitude $Q_T$ and spectral
index $n_T$. A neutrino species with a mass greater than
an eV affects structure formation,
so the number $N_\nu$ of light (meaning
$m_\nu\lesssim1$~eV) neutrinos is another cosmological parameter of
importance.  The ionization history of the Universe is also
certainly related to the evolution of structure in the Universe.

In this paper, we estimate how well cosmological parameters can
be determined from a CMB temperature map.  
Since the initial detection of temperature anisotropies in
the cosmic microwave background (CMB) by the COBE satellite
\cite{dmrorig}, over a dozen other balloon-borne and
ground-based experiments have announced anisotropy detections
on smaller angular scales \cite{wss}.  With the existence
of anisotropies now firmly established,
sights are shifting to accurate
determination of the CMB power spectrum over a wide
range of angular scales.  Several technological
advances, including improved amplifiers,
interferometry, and long-duration balloon flights, hold
great promise for high-precision measurements. Ultimately,
a satellite with sub-degree angular resolution will provide
a detailed map of the entire microwave sky in multiple frequency
bands \cite{cobe2}.

A detailed map of the cosmic microwave background can
potentially provide a wealth of information on the values of
cosmological parameters.  Roughly speaking, the amount
of information in a map is proportional to the number of pixels
on the sky, and this is inversely proportional to the square
of the beam width.  Thus, a map with a beam width of
0.5$^\circ$ will contain over 100 times as much information as
COBE, which had a beam width of order 7$^\circ$, and an
0.1$^\circ$-resolution experiment would have, roughly speaking,
$10^4$ times as much information!
It should be no surprise, therefore, that a map with good
angular resolution should be able to determine many more
cosmological parameters than COBE, which really only constrains
the normalization of the CMB power spectrum and the effective
CMB spectral index at large angular scales.

We consider
an experiment which maps a given fraction of the sky with a
given angular resolution and a given level of pixel noise.  We
use a covariance-matrix approach to evaluate the standard errors
which would arise by fitting the power spectrum obtained in this
experiment to all the unknown cosmological parameters.  We
display results for a range of realistic values for the fraction
of sky covered, level of pixel noise, and angular resolution.  Our
results are quite promising:  With minimal assumptions,
realistic satellite experiments could potentially determine
$\Omega$, $\Lambda$, and the inflationary observables to far
greater precision than any traditional measurements.
Furthermore, the information provided on other parameters will
be competitive with (and with additional reasonable assumptions,
superior to) current probes.  Although we focus here only on
models with primordial adiabatic perturbations, we are confident
that if the perturbations turn out to be
isocurvature, it will be evident in the temperature maps (and
perhaps also in polarization maps, spectral distortions, and
non-gaussian temperature distributions), and that similar results on
parameter determination will apply.  Indeed, recent calculations
of the CMB power spectrum in defect models \cite{davesworld}\
and in isocurvature models \cite{Peebles}\ suggest that such
models should be clearly distinguishable from the adiabatic case.
Although we have
satellite mapping experiments in mind, our results can also be
applied to ground or balloon experiments, or to the combined
results of several complementary measurements.

An important issue facing any likelihood
analysis is the choice of the space of models considered.
Here we consider models with primordial adiabatic
perturbations.  Our space of models allows a
cosmological constant, an open (or closed) Universe, tensor
modes (with a free spectral index), variations in the baryon
density and Hubble constant, tilted primordial spectra, and
primordial spectra that deviate from pure power laws.  We assume
that the dark matter is cold; however, the CMB power
spectrum is only slightly altered in mixed and hot dark-matter
models \cite{mdm}, and we allow the number of massless neutrinos
to vary. Therefore, our conclusions on parameter determination will
be virtually independent of the fraction of hot dark matter.

In the following Section, we describe our calculation of the
power spectrum. In Section III, we illustrate the effect of each
cosmological parameter that we consider on the CMB spectrum.  In
Section IV, we discuss the covariance matrix.  To illustrate, in Section
V, we present results for the standard errors to the parameters
that would be obtained assuming the true cosmological model is
standard CDM.  We also discuss how these results change if the
underlying model differs from the canonical standard-CDM model.
In Section VI, we discuss the validity of the covariance-matrix
approach to the analysis.  In Section VII, we make some
concluding remarks and discuss some future areas of
investigation.

\section{CALCULATION OF THE CMB SPECTRUM}

In many areas of astrophysics, it is difficult to make
detailed quantitative predictions as properties of complex
systems  depend on non-linear physics of poorly measured and
poorly understood phenomena.  Fortunately, the early Universe
was very simple and nearly uniform.  The density fluctuations
are all in the linear regime ($\delta \rho/\rho \sim 10^{-4}$)
and non-linear effects are unimportant.  Different groups using
different gauge choices and numerical algorithms  make very similar
predictions for CMB fluctuations for a given model.  This simple
linearity makes possible the detailed parameter determination
that we describe in this paper.

The CMB angular power spectrum $C(\theta)$ is defined as
\begin{equation}
     C(\theta)\equiv \left\langle 
     {\Delta T\over T_0}({\bf \hat m})
     {\Delta T\over T_0}({\bf \hat n})
     \right\rangle,\qquad
     {\bf \hat m}\cdot{\bf \hat n} = \cos\theta,
\label{Ctheta}
\end{equation}
where the angle brackets represent an ensemble average
over all angles and observer positions.
Here $\Delta T({\bf\hat n})/T_0$ is the fractional temperature
fluctuation in the direction $\bf\hat n$, and the mean CMB
temperature is $T_0=2.726 \pm 0.010 \,{\rm K}$ \cite{firas}.
This power spectrum is conveniently expressed in terms
of its multipole moments $C_\ell$, defined by
expanding the angular dependence in Legendre polynomials,
$P_\ell(x)$:
\begin{equation}
     C(\theta)= \sum_{\ell=2}^\infty {2\ell+1\over 4\pi} 
     C_\ell P_\ell(\cos\theta).
\label{Cldefinition}
\end{equation}
Given a model for structure formation, calculation of the
multipole moments is straightforward and is accomplished by
solution of the coupled system of Boltzmann equations for each
particle species (i.e., photons, baryons, massless and possibly
massive neutrinos, and cold dark matter) and Einstein equations for
the evolution of the metric perturbations.
The $\ell=1$ term is indistinguishable from the Doppler shift
due to proper motion with respect to the microwave background
rest frame and is conventionally ignored. For theories with
gaussian initial perturbations, the set of $C_\ell$ completely
specifies the statistical properties of the theory.
Since we can only observe from a single vantage point in the
Universe, the observed multipole moments $C_\ell^{\rm obs}$ will be
distributed about the mean value $C_\ell$ with a ``cosmic variance'' 
$\sigma_\ell \simeq \sqrt{2/(2\ell+1)}C_\ell$; no measurement
can defeat this variance.  Power-spectrum predictions and
measurements are traditionally plotted as $\ell(\ell+1) C_\ell$
versus $\ell$.

For the purposes of covariance-matrix evaluation, as well as for
likelihood maximization \cite{spergel,minimization}\ and Monte
Carlo analysis, it is useful to have an algorithm
for rapid evaluation of the CMB spectrum for a given set of
cosmological parameters.  We begin
with a semi-analytic solution of the coupled Boltzmann,
fluid, and Einstein equations
developed by Hu and Sugiyama \cite{hu} for
flat cold-dark-matter models, which we generalize to
accommodate an open Universe, a cosmological constant,
tensor modes, and reionization.  The code
is fast enough to enable likelihood analyses requiring
tens of thousands of power-spectrum evaluations.  We have
checked that our semi-analytic calculation agrees with the results
of a publicly available numerical code \cite{cosmics} for
several parameters.  Here we briefly describe our calculation.

The multipole moments are expressed as
\begin{equation}
     C_\ell=C_\ell^S+C_\ell^T,
\label{Cellsum}
\end{equation}
where $C_\ell^S$ is the contribution from scalar perturbations
and $C_\ell^T$ is the contribution from tensor modes.  The
scalar contribution is given by
\begin{equation}
     C_\ell^S = {2\over \pi} \int_0^\infty\,dk\, k^2
     |\Theta_\ell(\eta_0,k)|^2,
\label{Cellintegral}
\end{equation}
where $\eta_0$ is the conformal time today
(the conformal time $\eta=\int dt/a$
with $a$ the scale factor of the Universe normalized to
unity at matter-radiation equality). 
The contribution of wavenumber $k$
to the $\ell$th multipole moment is \cite{hu}
\begin{equation}
     \Theta_\ell(\eta_0,k) \simeq 
     [\Theta_0 + \Psi](k,\eta_*) j_l(k\eta_0- k\eta_*) 
     + \Theta_1(k,\eta_*) j_l'(k\eta_0 -k\eta_*) 
     + \int_{\eta_*}^{\eta_0}\,d\eta\,
     [\dot\Psi - \dot \Phi] j_l(k\eta_0 - k\eta),
\label{husugiyamatwelve}
\end{equation}
where $\Theta_0$ and $\Theta_1$ are the monopole and dipole
perturbations of the photon distribution function,
$\Phi$ and $\Psi$ are gravitational-potential
perturbations in the Newtonian gauge,
$j_l$ are spherical Bessel functions and
$j_l'$ their first derivatives, and a dot denotes derivative
with respect to conformal time.  
Here $\eta_*$ is the conformal time at decoupling.  (See
Ref. \cite{hu}\ for more details.)  
The third term in this expression gives the integrated
Sachs-Wolfe (ISW) effect: anisotropies are generated by
time variations in the gravitational potentials
along the line-of-sight path.
Analytic fits to the gravitational
potentials are given in Ref. \cite{hu}, as are WKB solutions
for the photon distributions in the tight-coupling regime,
$\hat\Theta_0$ and $\hat\Theta_1$.  At decoupling, photon
diffusion (Silk damping) damps photon
perturbations on small angular scales \cite{silk};  the perturbations
to the photon distribution functions are given by
$[\Theta_0+\Psi](\eta_*) = [\hat\Theta_0+\Psi](\eta_*){\cal D}(k)$,
where the mean damping factor is
\begin{equation}
     {\cal D}(k)=\int_0^{\eta_0}\, \dot \tau e^{-[k/k_D(\eta)]^2}
     d\eta.
\label{dampingfactor}
\end{equation}
Here $\dot\tau=x_e n_e\sigma_T a/a_0$ is the differential
optical depth for Thomson scattering, $n_e$ is the electron
density, $x_e$ is the ionization fraction, and $\sigma_T$
is the Thomson cross section.  The visibility function---the
combination $\dot\tau e^{-\tau}$ ---is the probability density
that a photon last scattered at given conformal time, and is
sharply peaked near the surface of last scatter; semi-analytic
fits are given in Ref. \cite{hu}.  As pointed out in
Ref. \cite{zh}, photon polarization must be included to obtain
the proper Silk-damping scale; the result is
\begin{equation}
     k_D^{-2}(\eta)={1\over6} \int_0^\eta\, d\eta{1\over
     \dot\tau} {R^2+16(1+R)/15 \over (1+R)^2},
\label{dampingscale}
\end{equation}
where
\begin{equation}
     R={3\rho_b\over 4\rho_\gamma}
      ={3\Omega_b a\over 4(1-f_\nu)\Omega_0}
\label{Requation}
\end{equation}
is the scale factor normalized to 3/4 at baryon-radiation
equality, with $\Omega_b$ the fraction of critical density in
baryons, $\Omega_0$ the fraction of critical density in
non-relativistic matter (baryons and cold dark matter), and
$f_\nu$ the fraction of the total radiation density
contributed by massless neutrinos.  Our numerical evaluation of
these expressions reproduces the power spectrum obtained from
Boltzmann codes to an accuracy of a few percent for standard
CDM.

Analytic approximations to the CMB anisotropy due to tensor modes
(gravity waves)
are given in Refs.~\cite{twl,wang}. The contribution to each
multipole moment of the CMB power spectrum is
\begin{equation}
C_\ell^{T} = 36\pi^2{(\ell+2)!\over(\ell-2)!}
       \int_0^\infty dk\, P_{T}(k)\,|F_\ell(k)|^2,
\label{tensorcl}
\end{equation}
where $P_{T}\propto k^{n_T+4}$ is the initial power spectrum of tensor
perturbations and $F_\ell$ is given by
\begin{equation}
F_\ell(k)\equiv k^{-3/2}\int_{\eta_*}^{\eta_0} d\eta\, \eta
           \left\{[1-w(\eta)] T\left({k\over k_{\rm eq}},\eta\right)
               {j_2(k\eta)\over(k\eta)^2}
                + w(\eta){j_1(k\eta)\over 3k\eta}\right\}
           {j_\ell(k\eta_0 - k\eta)\over(k\eta_0 - k\eta)^2},
\label{define_fl}
\end{equation}
with $k_{\rm eq}$ defined as the wavenumber of the mode which
enters the horizon at matter-radiation equality.
The fitting function $w(\eta)$ describes the evolution of the
gravity-wave mode function through the transition between the 
radiation-dominated and matter-dominated epochs, and $T(k,\eta)$
is a transfer function describing the evolution of the
tensor-mode amplitude. Good analytic fits to these two functions
are given by \cite{wang}
\begin{eqnarray}
w(\eta) &=& \exp\left(-0.2\eta^{0.55}\right)\\
T(y,\eta) &=& {\eta^2\over a}\left[ e^{-4y^4}
              (1 + 1.34y + 2.5y^2)^{1/2}
              + 1-e^{-4y^4}\right].
\label{fits}
\end{eqnarray}
These approximations match numerical results to
one percent well past $\ell=100$, where
the tensor contribution to the multipoles
drops to a small fraction of the scalar contribution.

Eqs.~(\ref{Cellintegral}) and (\ref{tensorcl})
are difficult to evaluate numerically because of the oscillatory
spherical Bessel functions in the integrand.  Asymptotic
expansions, a Bessel-function
cache, and various interpolation techniques
further speed evaluation of the integrals. We calculate
every 40th multipole (more for $\ell<100$)
and perform a cubic spline to recover the entire spectrum.

We consider models which are well-described by a power-law
spectrum of metric perturbations over the range of scales
affecting CMB anisotropies. This class includes all inflation
models. For the scalar perturbations, we also allow a
deviation from power-law behavior and parameterize the power
spectrum as \cite{kt}
\begin{equation}
     P(k)\propto \left({k\over k_S}\right)^{n_S+\alpha \ln(k/k_S)},
\label{pkscalar}
\end{equation}
where $k_S$ is the normalization scale at
which the power law index $n_S$ is defined. The parameter
$\alpha$ quantifies the deviation from the power law,
or the ``running'' of the spectral index. Realistic inflation
models can produce values of $\alpha$ large enough to 
change the multipole moments by as much as 5\%. For the
tensor spectrum, we assume a pure power-law spectrum
with spectral index $n_T$. In principle, $n_T$ can
run with scale as well, but because of the comparatively
small amount of information contained in the tensor
multipole moments, the CMB constraint
on the index $n_T$ is weak, and the running-index
effect for the tensor perturbations is negligible.

Extensions of this basic cosmological model are
incorporated through various fitting formulas.
In a cosmological-constant ($\Lambda$) Universe,
the gravitational potential $\Phi$ begins to vary at low
redshift when the Universe becomes cosmological-constant
dominated, and this leads to a contribution to the anisotropy at
large angles from the integrated Sachs-Wolfe (ISW) effect.  In a
flat Universe (that is, $\Omega_0+\Lambda=1$, where $\Lambda$ is
the cosmological constant in units of critical density), this is
approximated by multiplying the multipole moments by a
factor $[1+g(\Lambda)/\ell]$ \cite{kofman,kamspergel},
where
\begin{equation}
     g(\Lambda)=36\pi \int_0^{\eta_0} {1\over [F(0)]^2}\left({dF \over d\eta}
     \right)^2(\eta_0-\eta)\,d\eta,
\label{gequation}
\end{equation}
\begin{equation}
     F(\eta)={H\over a} \int\,{da/a_0 \over (Ha/a_0)^3}
\label{Feqn}
\end{equation}
is the time dependence of the potential, and $H=\dot a/a$ is the
Hubble parameter.  This approximation
slightly overestimates the lowest few multipole moments, but
this large-angle ISW effect is generally not a large fraction of the
total anisotropy, and the lowest
multipole moments have a limited statistical significance.  For
$\Lambda\lesssim 0.7$, $g(\Lambda)$ can be approximated by
\begin{equation}
    g(\Lambda)\simeq 0.637\left( {\Lambda \over 1-
    \Lambda} \right)^{0.817}.
\label{ccapprox}
\end{equation}
An additional effect of a cosmological constant is a shift in the
conformal distance to the surface of last scatter (even with
the mass density $\Omega_0 h^2$ held fixed), which we account
for by multiplying the current conformal time $\eta_0$ by the
correction factor $1 + 0.085 \ln(1-\Lambda)$ \cite{HS95}.

Generalization to an open Universe is somewhat more complicated
because several different effects contribute to the anisotropy
\cite{kamspergel}.  The
angular scale subtended by the horizon at the surface of last
scatter scales as $\Omega^{1/2}$ where
$\Omega=\Omega_0+\Lambda$ is the total density (in units
of critical density) of the Universe \cite{kamspergelsug}.
Therefore, the multipole moments in an open Universe are related
to those in a flat Universe approximately by $D_\ell(\Omega) \simeq
D_{\ell\Omega^{1/2}}(\Omega=1)$ with $D_\ell=\ell(\ell+1)
C_\ell$.  In other words, the CMB spectrum in an open Universe
resembles that in a flat Universe with the same matter density,
but shifted to smaller angular scales. A
large-angle ISW effect arises from the evolution of the
gravitational potentials, although
the function $g(\Omega)$ differs from that in
a cosmological-constant Universe \cite{kamspergel}.  In addition,
the lowest multipole moments probe scales 
comparable to or larger than the curvature
scale, so these moments are suppressed, due
heuristically to the exponential growth of volume in an open
Universe at large distances.  Finally, some ambiguity exists as to
the correct generalization of a power-law spectrum to an
open Universe.  Naive power laws of
volume, wavenumber, or eigenvalue of the Laplace operator
differ in an open Universe \cite{kamspergel}, as
do spectra predicted by various open-Universe inflationary
scenarios \cite{lythstewart}.  However,
these power laws differ only in their predictions
for the lowest multipole moments, which have little statistical
weight; for definiteness, we use the predictions of a specific
inflationary scenario \cite{kamratra}.  A good fit to these effects (for
$\Omega\gtrsim0.1$) is provided by multiplying the multipole moments
by 
\begin{equation}
     1+e^{-0.3\;\ell/\ell_{\rm curv}} {g(\Omega)\over \ell+1/2},
\label{marcsfit}
\end{equation}
where $\ell_{\rm curv}=\pi \sqrt{(1-\Omega)/\Omega}$ is the
multipole corresponding to the curvature scale of the Universe,
and
\begin{equation}
     g(\Omega)\simeq 4.5 \left( {1-\Omega \over \Omega}
     \right)^{0.817},
\label{openUgeqn}
\end{equation}
for $\Omega\gtrsim0.1$.

If the Universe has experienced significant reionization between
recombination and today, then a
fraction $1-e^{-\tau_{\rm reion}}$ of the CMB photons have scattered
since recombination, 
where $\tau_{\rm reion}$ is the optical depth to
the epoch of recombination. If the
Universe becomes reionized at a redshift $z_{\rm reion}$ with a
constant ionization fraction $x_e$, then the optical depth
is $\tau_{\rm reion}\simeq0.04\, \Omega_b h
\Omega^{-1/2} x_e [(1+z_{\rm reion})^{3/2}-1]$, where $h$ is the
Hubble parameter in units of 100~km~sec$^{-1}$~Mpc$^{-1}$.
The precise effects of reionization depend on the baryon
density, Hubble parameter, and the ionization history.  However,
as illustrated in Ref. \cite{kamspergelsug}\ (see Fig.~3
therein), the effects of reionization are fairly accurately
quantified solely in terms of $\tau_{\rm reion}$. Compton
scattering is an isotropizing process, so the multipole moments
on angular scales smaller than those subtended by the horizon at
the epoch of reionization are suppressed by a factor
$e^{-2\tau_{\rm reion}}$, while those on larger angular scales are
unaffected. We
interpolate between the asymptotic effects of reionization on
small and large angular scales by multiplying the
multipole moments by
\begin{equation}
     \exp\left[{-2\tau_{\rm reion} (\ell\eta_{\rm reion}/\eta_0)^2 \over 1 +
     (\ell\eta_{\rm reion}/\eta_0)^2} \right],
\label{reionizationdamping}
\end{equation}
where $\eta_{\rm reion}$ is the conformal time at reionization.
In addition, reionization also induces a broad Doppler
peak centered near $\ell\simeq\eta_0/\eta_{\rm reion}$
\cite{kaiser,huwhite}, but this secondary peak is shallow and we do
not include it in the power-spectrum calculation.

Between the surface of last scatter and the present,
several other physical processes, besides
reionization, produce new CMB fluctuations and smear out
primordial fluctuations \cite{tegmarkreview}: 
gravitational lensing lowers the amplitude of the spectral peaks
and fills in the valleys in the spectrum \cite{lensing}; 
the non-linear growth of structure
produces additional small-scale fluctuations \cite{reessciama};
the scattering of photons off of hot gas in clusters and superclusters
produces both thermal and non-thermal cosmic microwave fluctuations
\cite{szannrev,persi}; and second-order effects in a reionized Universe
also produce additional small-scale fluctuations \cite{vishniac,persi}.
These non-linear effects are relatively small and typically produce only
${\cal O}(\mu {\rm K})$
changes in the microwave multipoles.  However, they are systematic. If they
are not included in an analysis of a full-sky CMB map, they will lead
to systematic errors in parameter estimation.  We do not include these
effects in our sensitivity analyses as they are unlikely to alter the
size and shape of the error ellipsoid.  It will be important to
include these effects in any analysis of a future all-sky CMB map.

\section{COSMOLOGICAL PARAMETERS AND THE CMB SPECTRUM}

\begin{figure}[t]
\centerline{\psfig{figure=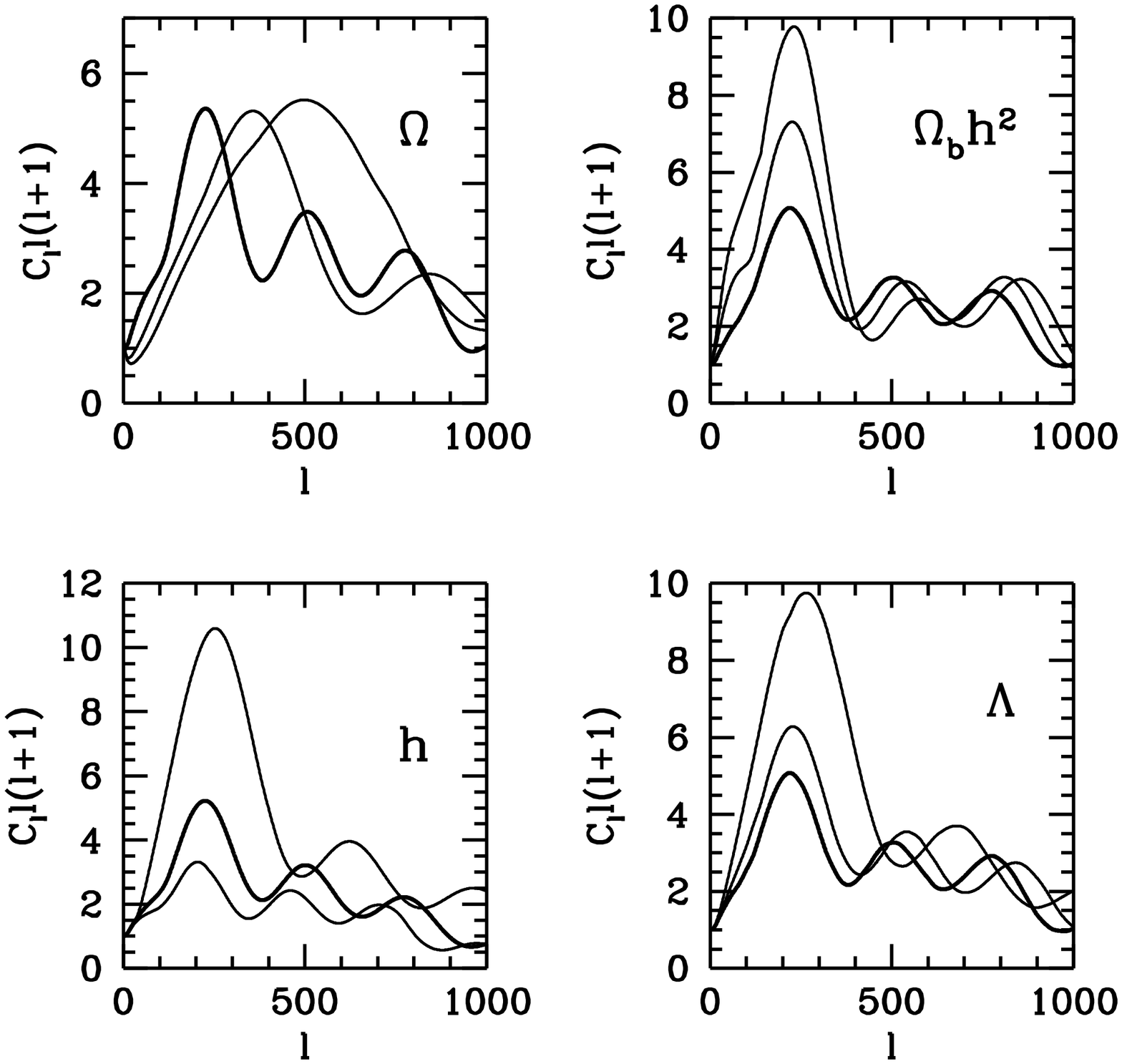,height=8in}}
\caption{Predicted multipole moments for standard CDM and
     variants.  The heavy curves in each graph are for a model
     with primordial adiabatic perturbations with $\Omega=1$,
     $\Lambda=0$, $n_S=1$, $\Omega_bh^2=0.01$, $h=0.5$,
     $\alpha=0$, and no tensor modes.  The graphs show the
     effects of varying $\Omega$, $\Lambda$, $h$, $\tau_{\rm
     reion}=0$, and $\Omega_b 
     h^2$ while holding all other parameters fixed.  In the
     $\Omega$ panel, from left to right, the solid curves are
     for $\Omega=1$, $\Omega=0.5$, and $\Omega=0.3$. 
     The curves in the $\Omega_b h^2$ panel are (from lower to
     upper) for $\Omega_bh^2=0.01$, $\Omega_bh^2=0.03$, and
     $\Omega_b h^2=0.05$.  In the $h$ panel, the heavy curves is
     for $h=0.5$, while the other two curves are for $h=0.3$
     (the upper light curve) and $h=0.7$ (the lower light
     curve).  The curves in the $\Lambda$ panel are for (from
     lower to upper) $\Lambda=0$, $\Lambda=0.3$, and
     $\Lambda=0.7$.}
\label{curvesfigure}
\end{figure}

The suite of cosmological models that we consider all
make broadly similar predictions for the CMB spectrum:
the fluctuations on large angular scales are nearly scale-invariant
and are primarily due to large-scale variations in the gravitational
potential at the surface
of last scatter, while on small scales the fluctuations are primarily due
to variations in the velocity and density of the baryon-photon fluid
at the surface of last scatter.  The details of the spectrum, however,
depend sensitively on properties of the Universe: its geometry, 
its size, the baryon density, the matter density,
and the shape of the primordial fluctuation spectrum.
In this Section, we discuss each
parameter that we consider and illustrate its most salient
effect on the CMB spectrum.  Fig.~\ref{curvesfigure}\
illustrates the following discussion.

The first Doppler peak occurs at the angular scale subtended
by the sonic horizon at the surface of last scatter.  Since
the photon energy density exceeds the baryon energy
density at that epoch, the sound speed of the Universe is
close to $c/\sqrt{3}$, so that the sonic horizon corresponds
to a nearly fixed physical scale.  The angular scale subtended
by this fixed physical scale will depend on the geometry
of the Universe.  
In an open Universe, the angular scale subtended by an object of
fixed diameter at fixed large redshift scales as $\Omega$.  On
the other hand, the causal horizon at last scatter is actually
$\Omega^{-1/2}$ times as large in an open Universe as it is in a
flat Universe.  Thus, to a first approximation, the
flat-Universe CMB spectrum is stretched by a factor
$\Omega^{1/2}$ to smaller angular scales in an open Universe.

Increasing the baryon density, $\Omega_b h^2$, reduces
the pressure at the surface of last scatter and therefore
increases the anisotropy at the surface of last scatter.
This reduction in pressure also lowers the sound speed of
the baryon-photon fluid, which alters the location and spacing
of the Doppler peaks.
Increasing the matter density, $\Omega_0 h^2$,
shifts matter-radiation equality to a higher redshift.  This
reduces the early-ISW contribution to the spectrum and lowers and narrows
the first Doppler peak.  If we knew that $\Lambda = 0$, then
the combination of these three effects (pressure, sound speed,
and redshift of matter-radiation equality) would be sufficient
to enable a determination of $\Omega_0, \Omega_b,$ and $h$
{}from the CMB spectrum.

The cosmological constant introduces a near degeneracy in parameter
determination.  Bond et al. \cite{confusion}\ stressed
that the CMB spectrum changed little if $\Lambda$ was
varied while $\Omega_0 h^2$ and $\Omega_b h^2$ were held fixed
in a flat Universe.  Changing $\Lambda$, however, does alter
the size of the Universe.  The conformal distance from the present
back to the surface of last scatter is smaller in a $\Lambda$-dominated
flat Universe than in a matter-dominated flat Universe.
Thus, increasing $\Lambda$ shifts the Doppler peak to larger angular
scales, the opposite effect of lower $\Omega_0$.  This
effect, along with the late-time ISW effect induced by $\Lambda$,
breaks the degeneracy and enables an independent
determination of all of the cosmological parameters
directly from an all-sky high-resolution CMB map.

The value of $N_\nu$, the effective number of noninteracting
relativistic degrees of freedom (in standard CDM, this is equal
to three for the three light-neutrino species), also shifts the
epoch of matter-radiation
equality and thus the height of the first Doppler peak as
discussed above.  In addition, if $N_\nu$ is changed, the
value of the anisotropic stress at early times---before the
Universe is fully matter dominated---is altered, and this
has a slight effect on the ISW contribution to the rise of the
first Doppler peak.

The tensor-mode contribution to the multipole moments simply
adds in quadrature with the scalar-mode contribution since there
is no phase correlation between them.  The amplitude of the
tensor modes is parameterized by $r=Q_T^2/Q_S^2$,
the ratio of the squares of the tensor and scalar contributions
to the quadrupole moment.\footnote{Note that this definition
differs from that in Ref.~\cite{us}.}\  The index $n_T$ is
defined so that the tensor-mode spectrum is roughly flat at
large angular scales for $n_T=0$;
it falls steeply near the rise to the first Doppler peak.
Thus, tensor modes may contribute to the anisotropy at large
scales, but they will have little or no effect on the structure
of the Doppler peaks.  Increasing the tensor spectral index,
$n_T$, increases the contribution at small angular scales relative
to those at larger angles.  

The overall normalization, $Q$, raises or lowers the spectrum
uniformly.  The effect of the scalar spectral index is similarly
simple: if $n_S$ is increased there is more power on small
scales and {\it vice versa}.  The effects of $\alpha$ are
obvious from Eq.~(\ref{pkscalar}).  Finally, the effects of
reionization have been discussed in the previous Section.

\section{ERROR ESTIMATES}

We consider an experiment which maps a fraction $f_{\rm sky}$
of the sky with a gaussian beam with full width at half maximum
$\theta_{\rm fwhm}$ and a pixel noise 
$\sigma_{\rm pix} = s/\sqrt{t_{\rm pix}}$, where $s$ is the detector
sensitivity and $t_{\rm pix}$ is the time spent observing each
$\theta_{\rm fwhm}\times\theta_{\rm fwhm}$ pixel. We adopt the
inverse weight per solid angle, 
$w^{-1}\equiv (\sigma_{\rm pix}\theta_{\rm fwhm}/T_0)^2$,
as a measure of noise that is pixel-size independent \cite{knox}.
Current state-of-the-art detectors achieve sensitivities of
$s=200\,\mu {\rm K}\,\sqrt{\rm sec}$, corresponding to an inverse
weight of $w^{-1}\simeq 2\times 10^{-15}$ for a one-year experiment.
Realistically, however, foregrounds and other systematic effects may
increase the effective noise level; conservatively, $w^{-1}$ will likely fall
in the range $(0.9-4)\,\times\,10^{-14}$.
Treating the pixel noise as gaussian and ignoring any
correlations between pixels, estimates of
$C_\ell$ can be approximated as normal distributions with
a standard error \cite{knox,hobson})
\begin{equation}
\sigma_\ell = \left[{2\over (2\ell +1)f_{\rm sky}}\right]^{1/2}
              \left[C_\ell + w^{-1}
	      e^{\ell^2\sigma_b^2}\right],
\label{variance}
\end{equation}
where $\sigma_b=7.42\times10^{-3}\, (\theta_{\rm
fwhm}/1^\circ)$.\footnote{We thank a referee for pointing out an
error in this equation in an earlier draft.}
Note that Eq. \ref{variance} applies only if the entire sky has
been mapped and then a fraction $1-f_{\rm sky}$ has been
subtracted.  On the other hand, if only a fraction $f_{\rm sky}$
of the sky is mapped, then the integration time per pixel
increases by a factor of $f_{\rm sky}^{-1}$, and $w^{-1}$ should
be replaced by $w^{-1}f_{\rm sky}$ \cite{hobson}.

\begin{figure}[t]
\centerline{\psfig{figure=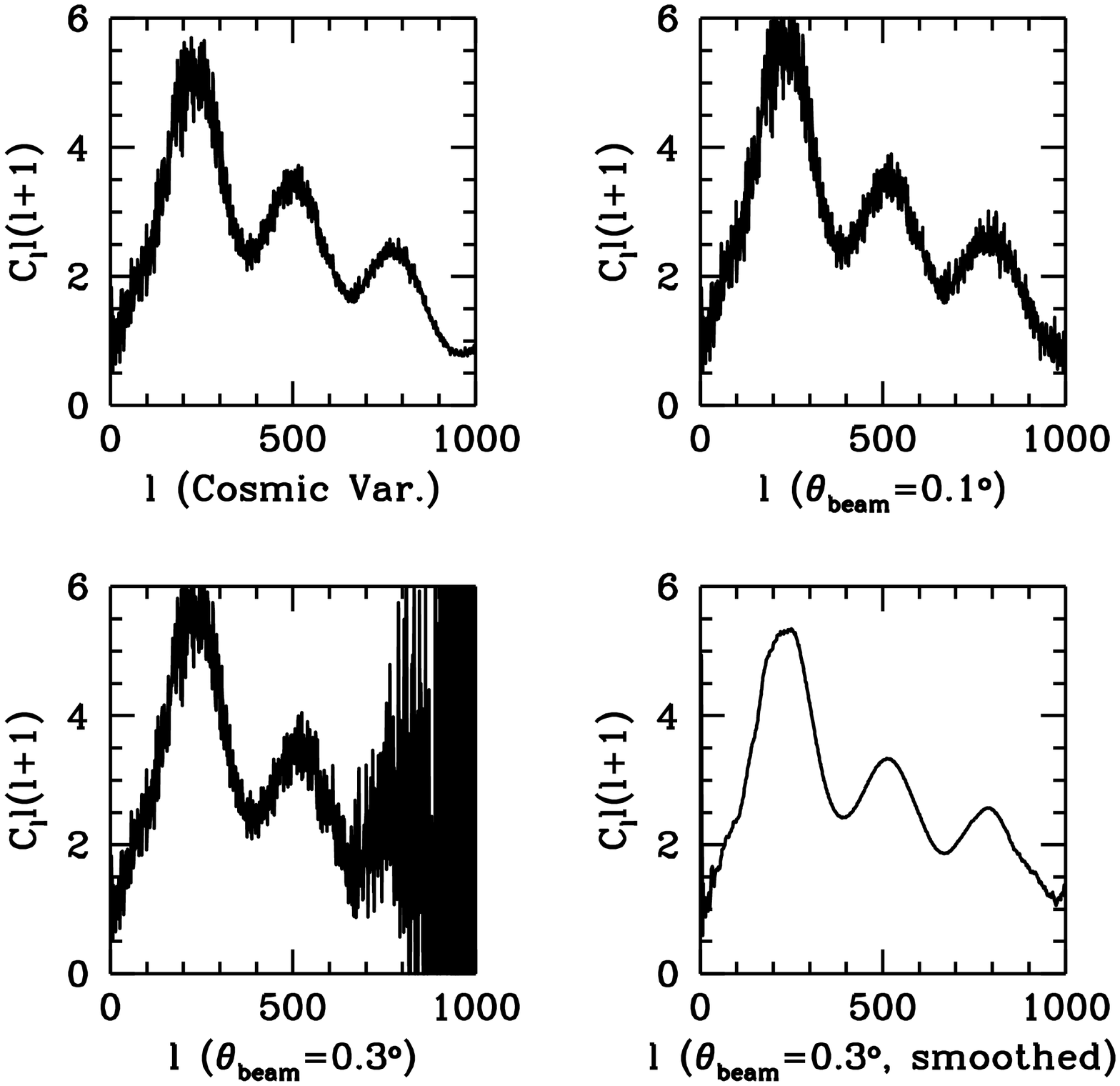,height=8in}}
\caption{Simulated data that might be obtained with a CMB
     mapping experiment, for beam sizes of $0.3^\circ$
     and $0.1^\circ$, and a noise level of $w^{-1}=2\times 10^{-15}$.}
\label{scatterfigure}
\end{figure}

In Fig.~\ref{scatterfigure}, we show simulated data that might
be obtained with a CMB mapping experiment, given an underlying
cosmological model of ``standard CDM'' (see the following section).
The ``Cosmic Variance'' panel illustrates the multipole moments
that would be measured by an ideal experiment (i.e., perfect
angular resolution and no pixel noise); the scatter is due only to
cosmic variance.  The top-right and bottom-left panels 
show multipole moments 
that might be measured by full-sky mapping experiments with a
realistic level of pixel noise and angular resolutions of
$0.1^\circ$ and $0.3^\circ$, respectively. The cosmic variance 
slightly increases the errors at lower $\ell$, 
while the finite beam width is evident in the increased
noise at $(\ell/700)\gtrsim(\theta_{\rm fwhm}/0.3^\circ)^{-1}$
in the lower-left plot. The lower-right panel
shows the moments from the lower-left panel after 
the total signal is smoothed
with a gaussian window of width $\ell/20$.  This
illustrates that although the individual moments may be
quite noisy, an experiment with a beam width of $0.3^\circ$
can still use the information in the location and shape of the third
peak in parameter estimation.  An experiment with this size beam
can extract useful information out to $\ell \sim 900$, although
it can not accurately measure the individual values
of these high $\ell$ multipoles.  
The smoothing here is used for display and is
not the optimal approach for parameter estimation.

We now wish to determine the precision with which a given CMB
temperature map will be able to determine the various cosmological
parameters.  The answer to this question will depend not only on
the experimental arrangement, but also on the correct underlying
cosmological parameters which we seek to determine.
For any given set of cosmological parameters,
${\bf s}= \{\Omega,\Omega_b h^2, h, \Lambda, n_S, r, n_T,
\alpha, \tau_{\rm reion}, Q, N_\nu\}$, the multipole moments,
$C_\ell({\bf s})$, can be calculated as described above. 
Suppose that the true parameters which describe the Universe are
${\bf s_0}$.  If the probability for observing each multipole
moment, $C_\ell^{\rm obs}$, is nearly a gaussian centered at
$C_\ell$ with standard error $\sigma_\ell$, and $\theta_{\rm fwhm}\ll
1$ so that the largest multipole moments sampled are $\ell \gg
1$, then the probability distribution for observing a CMB power
spectrum which is best fit by the parameters ${\bf s}$ is
\cite{us,gould,tegmarkreview}
\begin{equation}
     P({\bf s}) \propto \exp\left[ -{1\over 2}({\bf s}-{\bf s}_0)
                     \cdot [\alpha] \cdot({\bf s}-{\bf s}_0)\right]
\label{likelihood}
\end{equation}
where the curvature matrix $[\alpha]$ is given approximately by
\begin{equation}
\alpha_{ij} = \sum_\ell {1\over\sigma_\ell^2}
              \left[{\partial C_\ell({\bf s}_0)\over\partial s_i}
                    {\partial C_\ell({\bf s}_0)\over\partial s_j}\right].
\label{curvature}
\end{equation}
As discussed in Ref. \cite{us}, the covariance matrix $[{\cal
C}] = [\alpha]^{-1}$ gives an estimate of the standard errors that
would be obtained from a maximum-likelihood fit to data: the
standard error in measuring the parameter $s_i$ (obtained by
integrating over all the other parameters) is approximately
${\cal C}_{ii}^{1/2}$.  Prior information about the values of
some of the parameters---from other observations or by
assumption---is easily included.  In the simplest case, if some
of the parameters are known, then 
the covariance matrix for the others is determined by inverting
the submatrix of the undetermined parameters. For example, if
all parameters are fixed except for $s_i$, the standard error in $s_i$
is simply $\alpha_{ii}^{-1/2}$.

Previous authors have investigated the sensitivity of a given
experimental configuration to some small subset of the
parameters we investigate here.  For example, Knox investigated
the sensitivity of mapping experiments to the inflationary
parameters, $n_S$, $n_T$, and $r$, but assumed all other
parameters (including $\Omega_b$ and $h$) were known \cite{knox}.
Similarly, Hinshaw, Bennett, and Kogut investigated the sensitivity to
$\Omega_b$ assuming all other parameters were fixed
\cite{hinshaw}.  These were
Monte Carlo studies which mapped the peak of the likelihood
function.  Another technique is to repeatedly simulate an
experimental measurement of a given underlying theory,
maximize the likelihood in each case and see how well the
underlying parameters are reproduced \cite{spergel}.  Such
calculations require numerous evaluations of the CMB spectrum,
so the results have been limited to a small range of
experimental configurations.  If any of these analyses are limited
to a small subset of
cosmological parameters, they do not investigate the possible
correlation with other undetermined parameters and will
therefore overestimate the capability of the experiment to
measure the parameters under consideration.

The covariance-matrix approach has the advantage that numerous
experimental configurations and correlations between all the
unknown cosmological parameters can be investigated with minimal
computational effort.  For example, if there are $N$
undetermined parameters, then we need only $N+1$ evaluations
of the $C_\ell$'s to calculate the partial derivatives in
Eq.~(\ref{likelihood}).  Once these are evaluated, the curvature
matrix for any combination of $w^{-1}$ and $\theta_{\rm fwhm}$
for $f_{\rm sky}=1$ can be obtained trivially.  The results are
generalized to $f_{\rm sky}<1$ by multiplying the results for the curvature
matrix by $f_{\rm sky}^{-1}$ [c.f., Eqs.~(\ref{variance})\ and
(\ref{curvature})].
Furthermore, the covariance matrix includes
all correlations between parameters.  Therefore, our results
reproduce and generalize those in
Refs. \cite{knox,hinshaw,spergel}, and we comment on this
further below.

\section{COVARIANCE-MATRIX RESULTS}

As discussed above, the sensitivity of a CMB map to cosmological
parameters will depend not only on the experiment, but also on
the underlying parameters themselves.  For illustration,
we show results for a range of experimental parameters under the
assumption that the underlying cosmological parameters
take on the ``standard-CDM'' values, ${\bf
s_0}=\{1,0.01,0.5,0,1,0,0,0,0,Q_{\rm COBE},3\}$, where $Q_{\rm
COBE}=20\, \mu K$ is the COBE normalization \cite{cobenorm}.
(It assumes a Harrison-Zeldovich primordial spectrum, no
tensor modes, no cosmological constant, a flat Universe,
and the central big-bang nucleosynthesis value for the
baryon-to-photon ratio.)  After presenting results for this
assumed cosmological model, we briefly discuss how the results
will be altered for different cosmological models.

With the eleven undetermined cosmological parameters we survey
here---some of which are better determined by experiment than
others---there are an
endless number of combinations that could conceivably be
investigated.  Instead of running through all possible
permutations, we present results for the standard errors that can be
obtained with two extreme sets of assumptions.  First, we
consider the case where none of the parameters are known.  Then
we consider the results under the most optimistic assumption
that all of the other parameters, except the normalization (which
will never be determined more accurately by any other
observations), are fixed.  Realistically, prior information on
some of the parameters will be available, so the standard errors
will fall between these two extremes.

\begin{figure}[t]
\centerline{\psfig{figure=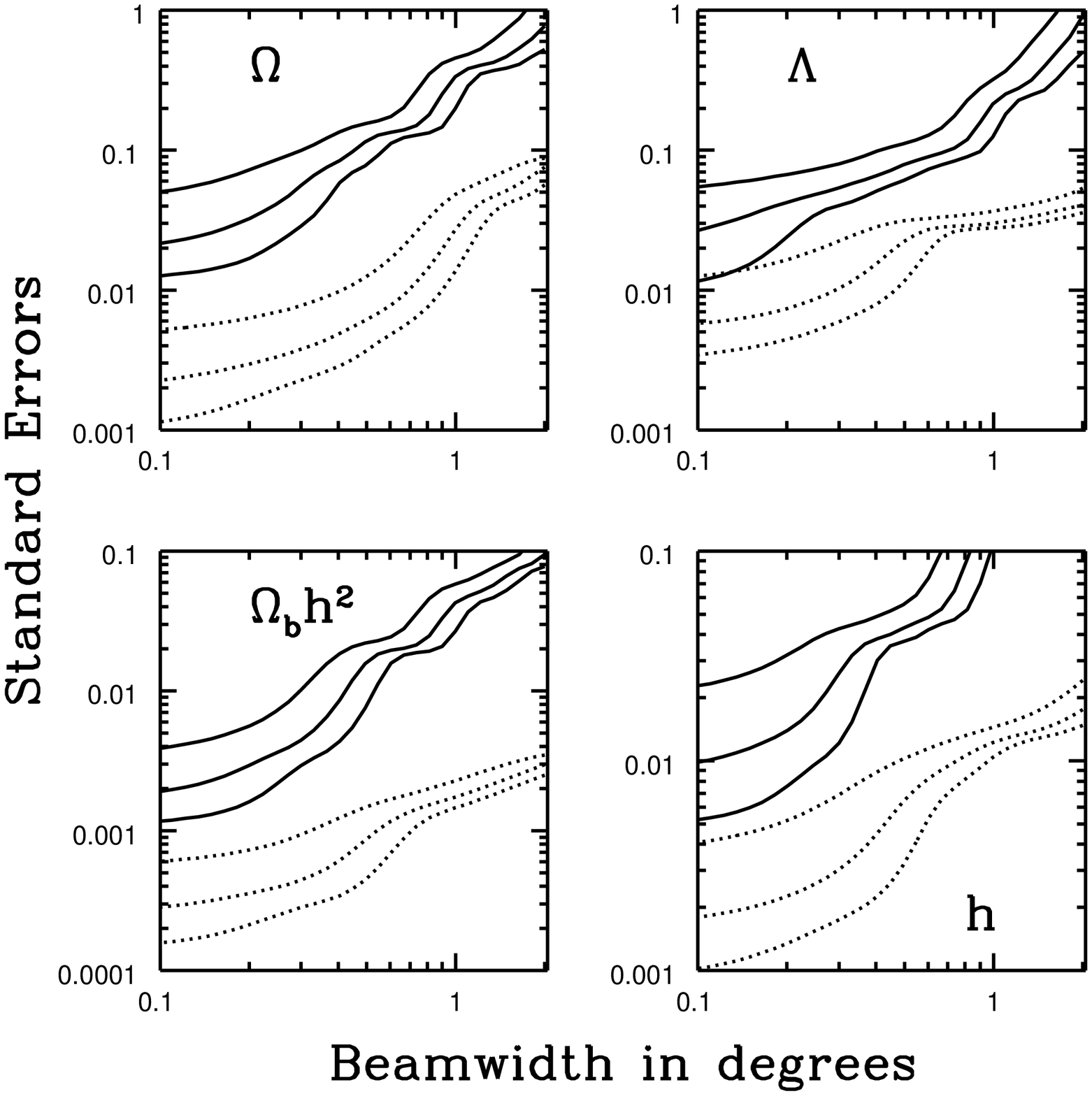,height=8in}}
\caption{The standard errors for $\Omega$, $\Lambda$, $\Omega_b
     h^2$, and $h$ that can be obtained with a 
     full-sky mapping experiment as a function of the beam width
     $\theta_{\rm fwhm}$ for noise levels $w^{-1}=2\times10^{-15}$,
     $9\times10^{-15}$, and $4\times10^{-14}$ (from lower to
     upper curves).  The underlying model is ``standard CDM.''
     The solid curves are the sensitivities attainable with no
     prior assumptions about the values of any of the other
     cosmological parameters.  The dotted curves are the
     sensitivities that would be attainable assuming that all
     other cosmological parameters, except the normalization
     ($Q$), were fixed.  The results for 
     a mapping experiment which covers only a fraction $f_{\rm
     sky}$ of the sky can be obtained by scaling by $f_{\rm
     sky}^{-1/2}$ [c.f., Eq.~(20)].}
\label{resultsone}
\end{figure}

\begin{figure}[t]
\centerline{\psfig{figure=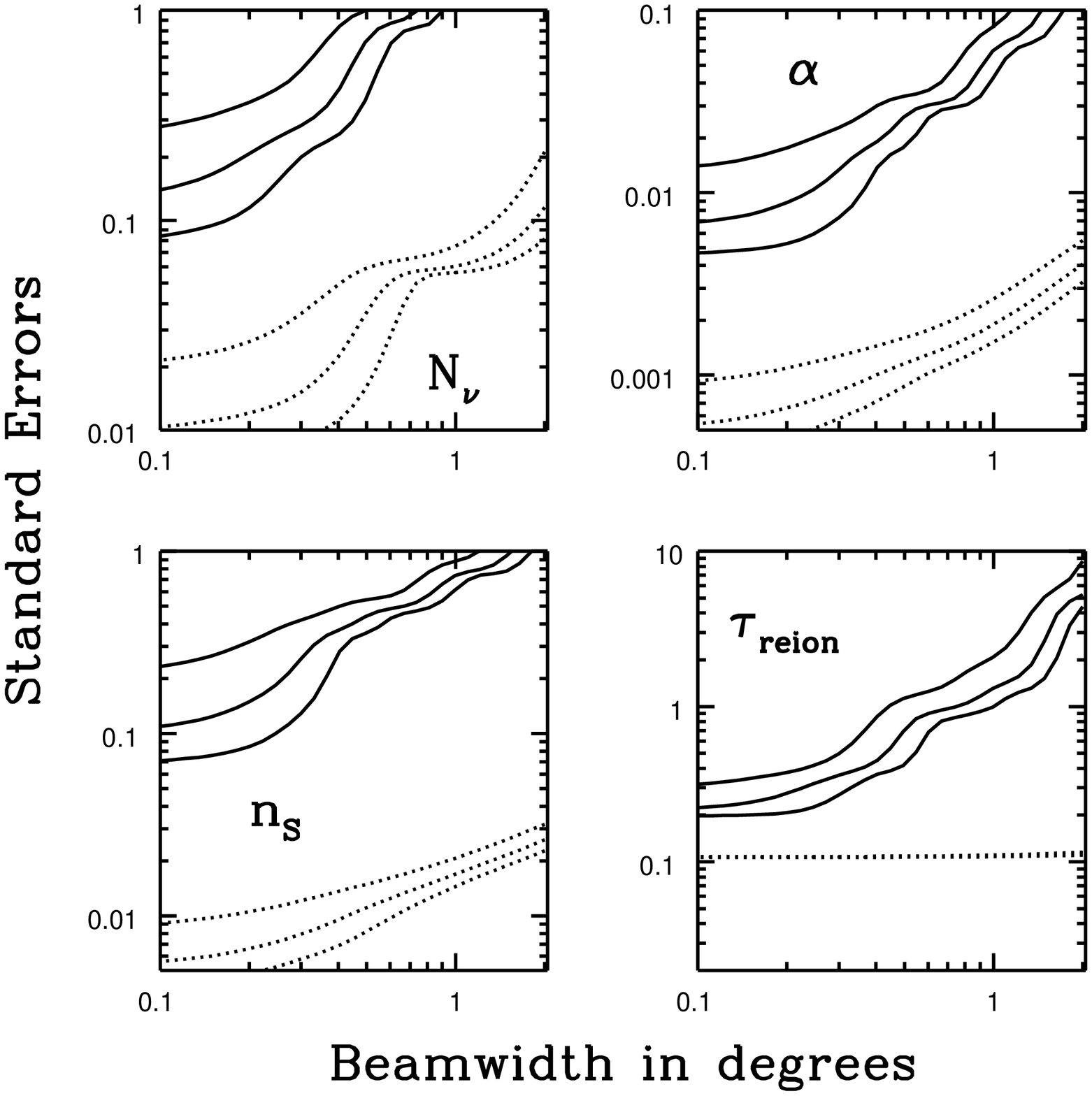,height=8in}}
\caption{Like Fig. 3, but for $\alpha$, $N_\nu$, $\tau_{\rm
     reion}$, and $n_S$.}
\label{resultstwo}
\end{figure}

Figs.~\ref{resultsone}\ and \ref{resultstwo}\ show the standard
errors for various
parameters that can be obtained with a full-sky mapping
experiment as a function of the beam width $\theta_{\rm fwhm}$
for noise levels $w^{-1}=2\times10^{-15}$, $9\times10^{-15}$,
and $4\times10^{-14}$ (from lower to upper curves).  The
underlying model is ``standard CDM.''  The solid curves are the
sensitivities attainable with no prior assumptions about the
values of any of the other cosmological parameters.  The dotted
curves are the sensitivities that would be attainable assuming
that all other cosmological parameters, except the normalization
($Q$), were fixed.  The analogous results for a mapping
experiment which covers only a fraction $f_{\rm sky}$ of the sky
can be obtained by scaling by $f_{\rm sky}^{-1/2}$ [c.f.,
Eq.~(\ref{variance})].

\subsection{The Total Density and Cosmological Constant}

The results for $\Omega$ were discussed in Ref.~\cite{us}.  {}From
the $\Omega$ panel in Fig.~\ref{resultsone}, it should be
clear that a CMB mapping experiment with sub-degree resolution
could potentially determine $\Omega$ to better than 10\% with
minimal assumptions, and perhaps better than 1\% with prior
information on other cosmological parameters.  This would be far
more precise than any conventional measurement of $\Omega$.
Furthermore, unlike mass inventories which measure only
the matter density $\Omega_0$, this measurement includes the
contribution to the density from a cosmological constant
(i.e., vacuum energy) and therefore directly probes the geometry
of the Universe.  This determination follows from the
angular location of the first Doppler peak.  Therefore, our
results show that if the Doppler peak is found to be at
$\ell\simeq200$, it will suggest a value of $\Omega=1$ to within
a few percent of unity.  This result will be {\it in}dependent
of the values of other cosmological parameters and will
therefore be the most precise test for the flatness of the
Universe and thus a direct test of the inflationary
hypothesis.  
Numerical calculations suggest that the
effect of geometry on the CMB spectrum may be slightly more
dramatic than indicated by our semi-analytic
algorithm.\footnote{We thank a referee for pointing this out.}
If so, our final results on the sensitivity to $\Omega$ are a
conservative estimate.

The sensitivity to $\Lambda$ is similar.  Currently,
the strongest bounds to the cosmological constant come from
gravitational-lensing statistics \cite{gravlenses}\ which only
constrain $\Lambda$ to be less than 0.5.  Measurement of the
deceleration parameter, $q_0=\Omega_0/2 - \Lambda$,
could provide some information on $\Lambda$, but the
measurements are tricky, and the result will depend
on the matter density.  On the other hand, a CMB mapping
experiment should provide a measurement of Lambda to
better than $\pm 0.1$, which will easily distinguish between a
$\Lambda$-dominated Universe and either
an open or flat matter-dominated Universe.

\subsection{The Baryon Density and Hubble Parameter}

The current range for the baryon-to-photon ratio allowed by
big-bang nucleosynthesis (BBN) is $0.0075 \lesssim \Omega_b h^2 \lesssim
0.024$ \cite{copi}.  This gives $\Omega_b\lesssim0.1$ for the range
of acceptable values of $h$, which implies that if $\Omega=1$, as
suggested by inflationary theory (or even if $\Omega\gtrsim0.3$ as
suggested by cluster dynamics), then the bulk of the mass in the
Universe must be nonbaryonic.  On the other hand, x-ray--cluster
measurements might be suggesting that the observed baryon
density is too high to be consistent with BBN \cite{felten};
this becomes especially intriguing
given the recent measurement of a large primordial
deuterium abundance in quasar absorption spectra \cite{hogan}.
The range in the BBN prediction can be traced primarily to
uncertainties in the primordial elemental abundances.
There is, of course, also some question as to whether the
x-ray--cluster
measurements actually probe the universal baryon density.
Clearly, it would be desirable to have an independent
measurement of $\Omega_b h^2$.  The $\Omega_b h^2$ panel in
Fig.~\ref{resultsone}\ shows that the CMB should provide such
complementary information.  The implications of CMB maps for the
baryon density depend quite sensitively on the experiment.  As
long as $\theta_{\rm fwhm}\lesssim0.5$, the CMB should (with minimal
assumptions) at least be
able to rule out a baryon-dominated Universe ($\Omega_b\gtrsim0.3$)
and therefore confirm the predictions of BBN.  With angular
resolutions that approach $0.1^\circ$ (which might be
achievable, for example, with a ground-based interferometry map
\cite{CIS}\ to complement a satellite map), a CMB map would
provide limits to the baryon-to-photon ratio that were
competitive with BBN.  Furthermore, if other parameters can be
fixed, the CMB might be able to restrict $\Omega_b h^2$ to a
small fraction of the range currently allowed by BBN.

Current state-of-the-art measurements of the Hubble parameter
approach precisions of roughly 10\%, and due to systematic
uncertainties in the distance ladder, it is unlikely that any
determinations in the foreseeable future will be able to improve
upon this result.  The panel for $h$ in Fig.~\ref{resultsone}\ 
shows that, even with minimal assumptions, a mapping experiment
with angular resolution better than $0.5^\circ$ will provide a
competitive measurement; with additional assumptions,
a much more precise determination is possible.  It should also
be noted that the CMB provides a measurement of the Hubble
parameter which is entirely independent of the distance ladder
or any cosmological distance determination.

As a technical aside, we mention that in calculating the curvature
matrix, Eq.~(\ref{curvature}), we choose $\Omega_0 h^2$ as an
independent parameter instead of $h$, and then transform the
curvature matrix back to the displayed parameters. The reason
for this choice is that the power spectrum depends on $h$ only
indirectly through the quantities $\Omega_0 h^2$ and $\Omega_b h^2$,
and the linear approximation to the change in the spectrum in
Eq.~(\ref{curvature}) is more accurate for the parameter $\Omega_0 h^2$.
This parameter choice also explicitly accounts for the
approximate degeneracy between models with the same value of
$\Omega_0 h^2$ but differing $\Lambda$ \cite{confusion}.

\subsection{Reionization}

As discussed above, the effects
of reionization can be quantified, to a first approximation, by
$\tau_{\rm reion}$, the optical depth to the surface of last scatter,
and there are several arguments which suggest
$\tau_{\rm reion}\lesssim1$ \cite{kamspergelsug}.  First of all,
significant reionization
would lead to anisotropies on arcminute scales due to the Vishniac
effect \cite{vishniac}, or to spectral (Compton-$y$)
distortions of the CMB \cite{spectral}.  Order-of-magnitude
estimates for the values of $\tau_{\rm reion}$ expected in adiabatic models
based on Press-Schechter estimates of the fraction of mass in
collapsed objects as a function of redshift suggest that
$\tau_{\rm reion}$
is probably less than unity \cite{kamspergelsug,tegmark}.  Moreover, the
numerous detections of anisotropy at the degree scale
\cite{wss}\ also show an absence of excessive
reionization.  Assuming complete reionization at a redshift
$z_{\rm reion}$, the optical depth with our standard-CDM values
is $\tau_{\rm reion}\simeq0.001\,z_{\rm reion}^{3/2}$, so
$\tau_{\rm reion}\lesssim1$ corresponds to $z_{\rm
reion}\lesssim100$.

The $\tau_{\rm reion}$ panel of Fig.~\ref{resultstwo}\
illustrates that,
with minimal assumptions, any map with sub-degree angular
resolution will probe the ionization history (i.e., $z_{\rm
reion}\lesssim1000$), and maps with resolutions better than
a half degree can restrict the optical depth to 0.5 or
less. While different ionization histories with the same
total optical depth can give different power spectra,
as long as the reionization is not too severe, simple
damping of the primary anisotropies is always the
dominant effect. The lower curves, assuming other
parameters are fixed except for $Q$, are flat because
at scales smaller than $2^\circ$, the effects of $\tau_{\rm reion}$
are precisely degenerate with a shift in $Q$. The lower curves
nearly coincide for the different noise levels because
all of the leverage in distinguishing $\tau_{\rm reion}$ comes
at low $\ell$ where the degeneracy with $Q$ is broken, and
at these scales the cosmic variance dominates the measurement
errors.

Although temperature maps alone may not provide a stringent
probe of the ionization history, polarization maps may
provide additional constraints \cite{minimization}.
The polarization produced at recombination is generally
small, but that produced during reionization can be much
larger.  Heuristically, the temperature anisotropy which is
damped by reionization goes into polarization.  Therefore, it is
likely that polarization maps will be able to better constrain
$\tau_{\rm reion}$ when used in conjunction with temperature
maps.

\subsection{Neutrinos}

We have also investigated the sensitivity of CMB anisotropies to
$N_\nu$, the effective number of neutrino degrees of freedom at
decoupling.  The number of
non-interacting relativistic species affects the CMB
spectrum by changing the time of matter-radiation equality,
although this cannot be distinguished from the same effect due to
changes in $h$, $\Omega_0$, and $\Lambda$.  However,
neutrinos (and other non-interacting degrees of
freedom which are relativistic at decoupling) free stream and
therefore have a unique effect on the growth of potential
perturbations.  This will be reflected in the
detailed shape of the CMB spectrum, especially from the
contribution of the early-time ISW effect.  In standard CDM,
there are the three light-neutrino species.  However, some
particle-physics models predict the existence of additional
very light particles which would exist in abundance in the
Universe.  Furthermore, if one of the light neutrinos has a mass
greater than an eV, as suggested by mixed dark-matter
models \cite{primack}\ and possibly by the Los Alamos experiment
\cite{losalamos}, then it would be nonrelativistic at decoupling
so the effective number of neutrinos measured by the CMB would
be $N_\nu<3$.\footnote{In such a case, the massive neutrino would
have additional effects on the CMB
\cite{mdm}.  Although we have not included these effects,
our analysis still probes variations in $N_\nu$, and our results
are suggestive of the sensitivity of CMB
anisotropies to a massive neutrino.}\  These limits would be
similar to limits on
the number of relativistic noninteracting species from BBN.
However, at the time of BBN, any particle with mass less than an
MeV would be relativistic, whereas at decoupling, only those
with masses less than an eV would be relativistic, so the
quantities probed by BBN and by the CMB are somewhat different.

The panel for $N_\nu$ in Fig.~\ref{resultstwo}\ shows the
sensitivity of CMB anisotropies to variations in the effective
number of non-interacting nonrelativistic species at decoupling.
When one takes into account systematic uncertainties in
primordial elemental abundances, BBN constrains the effective
number of relativistic (i.e., less than a few MeV) neutrino species to
be less than 3.9 \cite{copi}.  Fig. 1 illustrates that any
mapping experiment with angular resolution better than
$0.5^\circ$ should provide complementary information; if other
parameters can be determined or constrained, then the CMB has
the potential to provide a much more precise probe of
the number of light neutrinos at the decoupling epoch.

\subsection{Inflationary Observables}

We have also studied the precision with which the inflationary
observables, $n_S$, $n_T$, and $r$, can be probed.  Inflation
predicts relations between the scalar spectral index $n_S$, the
tensor spectral index $n_T$, and the ratio $r$ 
\cite{liddle}.  Therefore, precise measurement of these
parameters provides a test of inflationary cosmology and perhaps
probes the inflaton potential \cite{reconstruction}.

Knox \cite{knox}\ performed a Monte Carlo calculation to address
the question of how accurately CMB anisotropies can
measure the inflationary observables assuming all other
cosmological parameters were known.
Here, we generalize the results to a broader range of
pixel noises and beam widths and take into account the
uncertainties in all other cosmological parameters through the
covariance matrix.

\begin{figure}[t]
\centerline{\psfig{figure=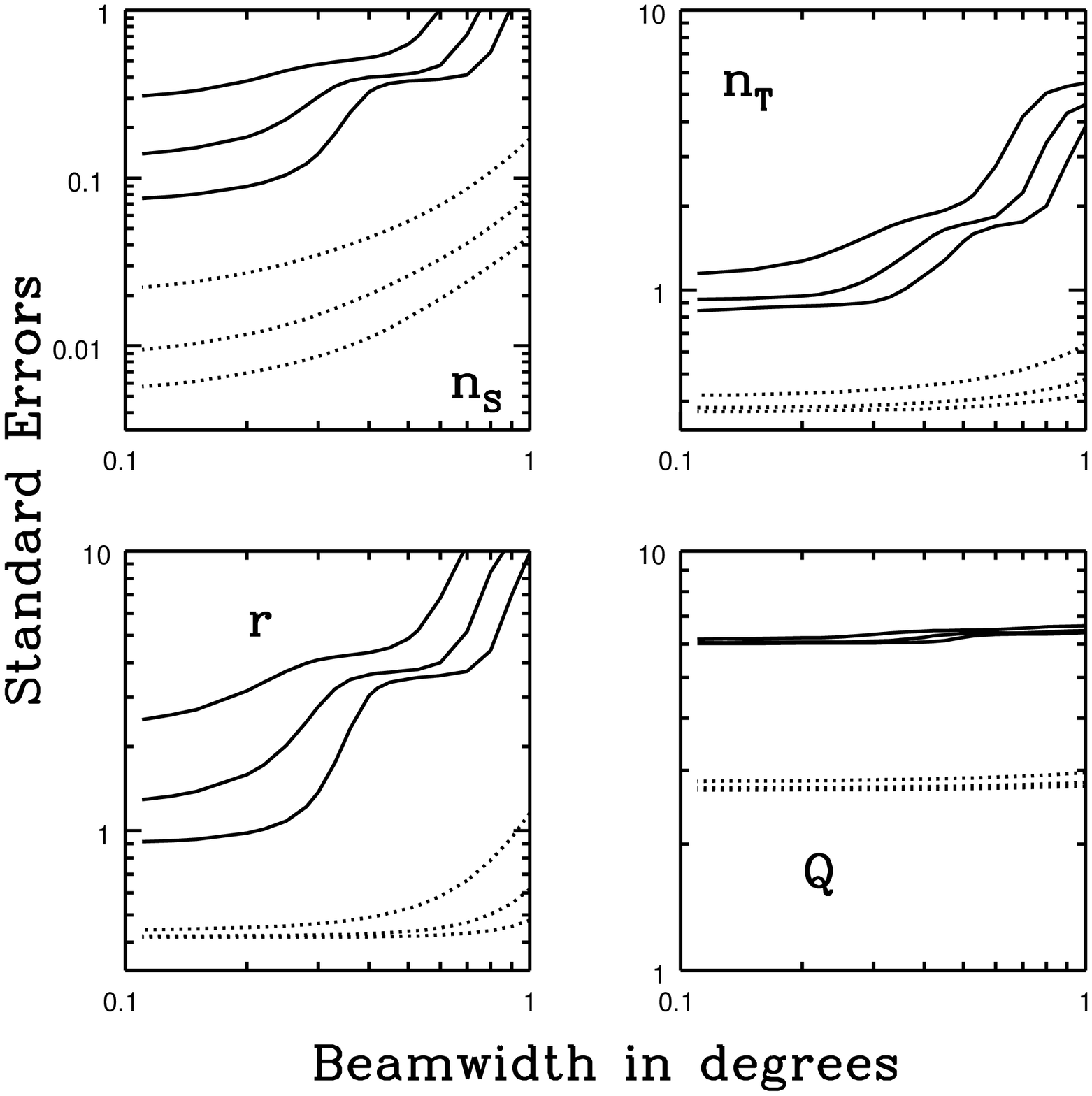,height=8in}}
\caption{The standard errors on the inflationary observables,
     $n_S$, $n_T$, $r=Q_T^2/Q_S^2$, and $Q$, that can be
     obtained with a
     full-sky mapping experiment as a function of the beam width
     $\theta_{\rm fwhm}$ for noise levels $w^{-1}=2\times10^{-15}$,
     $9\times10^{-15}$, and $4\times10^{-14}$ (from lower to
     upper curves).  The parameters of the underlying model are
     the ``standard=CDM'' values, except we have set $r=0.28$,
     $n_S=0.94$, and $n_T=-0.04$.
     The solid curves are the sensitivities attainable with no
     prior assumptions about the values of any of the other
     cosmological parameters.  The dotted curves are the
     standard errors that would be attainable by fitting to only
     these four inflationary observables and assuming all other
     cosmological parameters are known.  (Note that this differs
     from the dotted curves in Fig.~4.)
     The results for 
     a mapping experiment which covers only a fraction $f_{\rm
     sky}$ of the sky can be obtained by scaling by $f_{\rm
     sky}^{-1/2}$ [c.f., Eq.~(20)].}
\label{lloydfigure}
\end{figure}

In Fig.~\ref{lloydfigure}, we show the standard errors on the
inflationary observables that could be obtained with mapping
experiments with various levels of pixel noise and beam widths.
The parameters of the underlying model used here are the same
``standard-CDM'' parameters used in Figs.~\ref{resultsone}\ and
\ref{resultstwo}, except here we set $r=Q_T^2/Q_S^2=0.28$, $n_S=0.94$, and
$n_T=-0.04$.  We do so for two reasons:  First, the tensor
spectral index is unconstrained without a tensor contribution;
second, these parameters will facilitate comparison with the results of
Ref.~\cite{knox}.  The solid curves are the standard errors that
would be obtained with no assumptions about the values of these
or any other of the cosmological parameters.  The dotted curves
are the standard errors that would be attainable by fitting to
only these four inflationary observables and assuming all other
cosmological parameters are known.  (Note that this differs from
the dotted curves in Figs.~\ref{resultsone}\ and \ref{resultstwo}.) 

The dotted curves in Fig.~\ref{lloydfigure}\ with a beam width
of $0.33^\circ$ are in good numerical agreement with the results of
Ref.~\cite{knox}.  This
verifies that the covariance-matrix and Monte Carlo calculations
agree.  Next, note that unless the other cosmological parameters
can be determined (or are fixed by assumption), the results of
Ref.~\cite{knox}\ for the sensitivities of CMB anisotropy maps to
the inflationary observables are very optimistic.  In
particular, temperature maps will be unable to provide any
useful constraint to $r$ and $n_T$ (and it will be impossible to
reconstruct the inflaton potential) unless the other parameters
can be measured independently.  However, if the classical
cosmological parameters can be determined by other means (or
fixed by assumption), the dotted curves in
Fig.~\ref{lloydfigure}\ show that fairly precise information
about the inflaton potential will be attainable.  CMB
polarization maps may provide another avenue towards improved
determination of the inflationary observables
\cite{minimization,crittenden}.

The flatness of the dotted curves for $r$ and $n_T$ in
Fig.~\ref{lloydfigure}\ is due to the fact that the contribution
of the tensor modes to the CMB anisotropy drops rapidly on
angular scales smaller than roughly a degree.  The solid curves
decrease with $\theta_{\rm fwhm}$ because the other cosmological
parameters (e.g., $\Omega_b h^2$ and $h$) become determined with
much greater precision as the angular resolution is improved.

Of course, the precision with which the normalization of the
perturbation spectrum can be measured with CMB anisotropies
(even current COBE measurements) is---and will continue to
be---unrivaled by traditional cosmological observations.
Galactic surveys probe only the distribution of visible mass,
and the distribution of dark matter could be significantly
different (this is the notion of biasing).  
The dotted figures in the panel for $Q$ in
Fig.~\ref{lloydfigure}\ are the sensitivities that would be
obtained assuming all other parameters were known.  
This standard error would be slightly larger if there were no
tensor modes included, because as $r$ is increased (with
the overall normalization $Q$ held fixed), the scalar
contribution is decreased.  Therefore, tensor modes decrease the
anisotropy on smaller angular scales, the signal to noise is
smaller, and the sensitivity to $Q$ (and other parameters) is
slightly decreased.  The effect of variations in other
underlying-model parameters on our results is discussed further
below.

\subsection{What If The Underlying Model Is Different?}

Now we consider what might be expected if the underlying
theory differs from that assumed here.  Generally, the parameter
determination will be less precise in models in which there is
less cosmological anisotropy, as reflected in
Eq.~(\ref{variance}).

What happens if the normalization differs from the central COBE
value we have adopted here?  The normalization raises or lowers
all the multipole moments; therefore, from Eq.~(\ref{variance}),
the effect of replacing $Q$ with $Q'$ is equivalent to replacing
$w^{-1}$ with $w^{-1}(Q/Q')$.  In Figs.~\ref{resultsone},
\ref{resultstwo}, and \ref{lloydfigure}, the solid
curves, which are spread over values of $w^{-1}$ that differ by more
than an order of magnitude, are all relatively close.  On the
other hand, the uncertainty in the COBE normalization is ${\cal
O}(10\%)$.  Therefore, our results are insensitive to the
uncertainty in the normalization of the power spectrum.

If there is a significant contribution to the COBE-scale
anisotropy from tensor modes, then the normalization of the
scalar power spectrum is lower, the Doppler peaks will be lower,
and parameter determinations that depend on the Doppler-peak
structure will be diluted accordingly.  On the other hand,
the tensor spectral index, which is important for testing
inflationary models, will be better determined.

Similarly, reionization damps structure on Doppler-peak
angular scales, so if there is a significant amount of
reionization, then much of the information in the CMB will be
obscured. On the other hand, there are several indications
summarized above that
damping due to reionization is not dramatic.  In
Ref. \cite{us}, we displayed (in Fig. 2 therein) results for the
standard error in $\Omega$ for a model with $\tau_{\rm reion}=0.5$.
As expected, the standard error is larger (but by no more than a
factor of two) than in a model with no reionization.

If $\Lambda$ is non-zero, $h$ is small, or $\Omega_b h^2$
is large, then the signal-to-noise should increase and there
will be more information in the CMB.  If the scalar spectral
index is $n_S>1$, then the Doppler peaks will be higher, but in
the more likely case (that predicted by inflation), $n_S$ will
be slightly smaller than unity.  This would slightly decrease
the errors.

If $\Omega$ is less than unity, then the Doppler peak (and all
the information encoded therein) is shifted to smaller angular
scales.  Thus one might expect parameter determinations to
become less precise if $\Omega<1$.  By explicit
numerical calculation, we find that for $\Omega=0.5$ (with all
other parameters given by the ``standard-CDM'' values), our
estimates for the standard error for $\Omega$ is at the same level
as the estimate we obtained for $\Omega=1$.  Therefore,
even if $\Omega$ is as small as 0.3, our basic conclusions that
$\Omega$ can be determined to $\pm0.1$ with minimal assumptions
are valid. The sensitivity to some of the other parameters,
in particular $h$, is (not surprisingly) significantly degraded
in an open Universe.

\section{GAUSSIAN APPROXIMATION TO THE LIKELIHOOD FUNCTION}

It is an implicit assumption of the covariance-matrix analysis
that the likelihood function has an approximate gaussian form
within a sufficiently large neighborhood of the maximum-likelihood
point.  Eq.~(\ref{likelihood}) only approximates the likelihood
function in a sufficiently small neighborhood around the
maximum.  The detailed functional form of the likelihood
function is given in Ref.~\cite{knox}.
If the likelihood function fails to be sufficiently
gaussian near the maximum, then the covariance-matrix method
is not guaranteed to produce an estimate for the standard error,
and a more involved (Monte Carlo) analysis would be essential.
Therefore, in the following we indicate the applicability
of the gaussian assumption for the likelihood.

First, we note that our parameterization has the property
that the individual parameters are approximately independent.
This is suggested by direct examination of the eigenvectors of the
covariance matrix. This is also supported by preliminary Monte Carlo
results. Therefore it is simplest to examine the behavior of the
likelihood as a function of individual parameters in order to determine
if a parabolic approximation to $\ln({\cal L})$ (the
log-likelihood) is admissible.

\begin{figure}[t]
\centerline{\psfig{figure=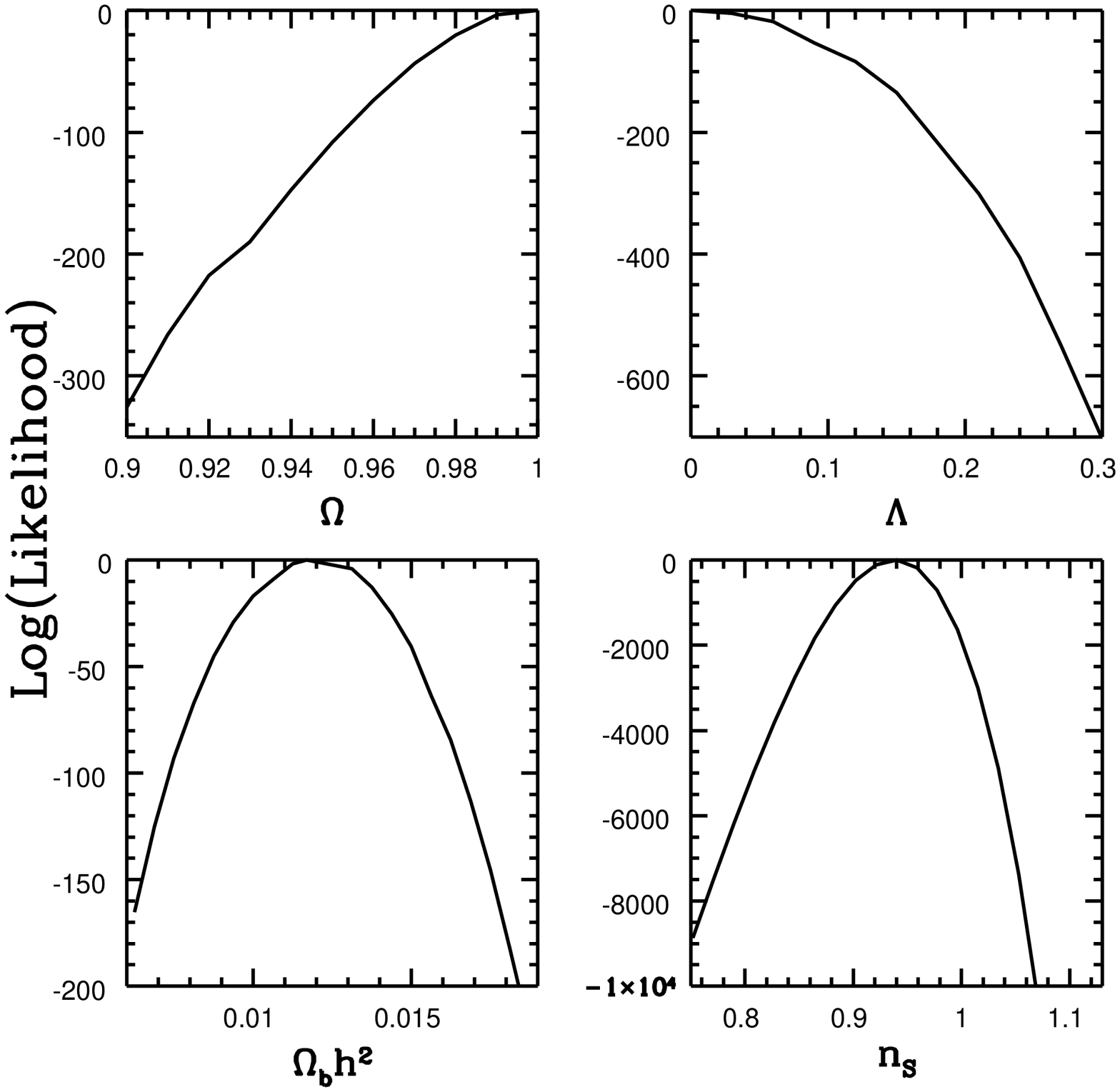,height=8in}}
\caption{Plots of the log-likelihood as a function of $\Omega$,
     $\Lambda$, $\Omega_b h^2$, and $n_S$ for the
     ``standard-CDM'' model with tensor modes (so $n_S=0.94$ and
     $r=0.28$).  Note that the log-likelihood looks parabolic.}
\label{gaussfigure}
\end{figure}

In Fig.~\ref{gaussfigure}, we display the dependence of
$\ln({\cal L})$ on several parameters of interest.  In this
example, we used $w^{-1}=9\times10^{-15}$ and $\theta_{\rm
fwhm}=0.25$.  As is clear from this Figure,
the functional forms are well fit by parabolic approximations,
within regions of size $\sim 3\sigma$ around the maximum point.
This is sufficient to apply the covariance-matrix analysis to the
determination of the standard errors, and our analysis above
is justified.

Although we have not done an exhaustive survey of the likelihood
contours in the eleven-dimensional parameter space,
Fig.~\ref{gaussfigure}\ also suggests that there are no
local maxima anywhere near the true maximum.  
Therefore, fitting routines will probably not be
troubled by local maxima.  This also suggests, then, that there
will be no degeneracy
between various cosmological models with a CMB map (in contrast
to the conclusions of Ref. \cite{confusion}), unless the models
are dramatically different.  In this event (which we consider
unlikely), one would then be forced to choose between two quite
different models, and it is probable that additional data would
determine which of the two models is correct.

\section{CONCLUSIONS}

We have used a covariance-matrix approach to
estimate the precision with which eleven cosmological parameters
of interest could be determined with a CMB temperature map.  We
used realistic
estimates for the pixel noise and angular resolution and
quantified the dependence on the assumptions about various
cosmological parameters that would go into the analysis.  The
most interesting result is for $\Omega$:  With only the
minimal assumption of primordial adiabatic perturbations,
proposed CMB satellite experiments 
\cite{cobe2}\ could potentially measure $\Omega$ to ${\cal
O}(5\%)$.  With prior information on the values of other
cosmological parameters possibly attainable in the
forthcoming years, $\Omega$ might be determined to better than
1\%.  This would provide an entirely new and independent
determination  of $\Omega$ and would be far more accurate
than the values given by any traditional cosmological
observations.  Furthermore, typical mass inventories yield only
the matter density.  Therefore, they tell us nothing about
the geometry of the Universe if the cosmological constant is
nonzero.  A generic prediction of inflation
is a flat Universe; therefore, locating the Doppler peak will provide a
crucial test of the inflationary hypothesis.

CMB temperature maps will also provide constraints on $\Lambda$
far more stringent than any current ones, and will provide a
unique probe of the inflationary observables.  Information on
the baryon density and Hubble constant will complement and perhaps
even improve upon current observations. Furthermore, although we have yet 
to include polarization maps in our error estimates, it is
likely that they will provide additional information, at 
least regarding ionization history.

We have attempted to display our results in a way that will be
useful for future CMB experimental design.  Although a satellite
mission offers the most promising prospect for making
a high-resolution CMB map, our error
estimates should also be applicable to complementary
balloon-borne or ground-based experiments which map a limited
region of the sky.  The estimates presented here can also be
used for a combination of complementary experiments.

Although we have been able to estimate the precision with which
CMB temperature maps will be able to determine cosmological
parameters, there is still much theoretical work that needs to
be done before such an analysis can realistically be carried
out.  To maximize the likelihood in a multidimensional parameter
space, repeated evaluation of the CMB spectrum for a broad range
of model parameters is needed.  Therefore, quick and
accurate calculations of the CMB anisotropy spectrum will be
crucial for the data analysis.  Several independent numerical
calculations of the CMB anisotropy spectrum now agree to roughly
1\%  \cite{COMBA}.  However, these calculations typically require
hours of workstation time per
spectrum and are therefore unsuitable for fitting data.
We have begun to extend recent analytic
approximations to the CMB spectra \cite{hu,seljak}
with the aim of creating a highly accurate and efficient
power spectrum code. Our current code
evaluates spectra in a matter of seconds on a
workstation, though our calculations are not yet as accurate as
the full numerical computations, except in a limited region of
parameter space.  It is likely, however, that the analytic
results can be generalized with sufficient accuracy.

The other necessary ingredient will be an efficient and reliable
likelihood-maximization routine.  Preliminary fits to simulated
data with a fairly simple likelihood-maximization algorithm
suggest that the cosmological parameters can indeed be
reproduced with the precision estimated here \cite{spergel}.  

In summary, our calculations indicate that CMB temperature maps
with good angular resolution can provide an unprecedented amount
of quantitative information on cosmological parameters.  These
maps will also inform us about the origin of structure in the
Universe and test ideas about the earliest Universe.  We hope
that these results provide additional impetus for experimental
efforts in this direction.

\acknowledgments

We would like to thank Scott Dodelson, Lloyd Knox, Michael Turner,
and members of the MAP collaboration for helpful discussions,
and George Smoot and an anonymous referee for useful comments.
Martin White graciously provided Boltzmann-code
power spectra for benchmarking our code.
This work was supported in part by the D.O.E. under contracts
DEFG02-92-ER 40699 at Columbia University
and DEFG02-85-ER 40231 at Syracuse University, by the Harvard
Society of Fellows, by the NSF under contract ASC 93-18185
(GC3 collaboration) at Princeton University,
by NASA under contract NAG5-3091 at Columbia University and
NAGW-2448 and under the MAP  Mission Concept Study Proposal at
Princeton University.  Portions of this work were completed
at the Aspen Center for Physics.

\vfil\eject

\end{document}